\documentclass[onecolumn,showpacs]{revtex4}

\usepackage{graphicx}
\usepackage{dcolumn}
\usepackage{bm}

\input epsf

\begin{document}

\title{\Large The effect of Pressure in Higher Dimensional Quasi-Spherical Gravitational Collapse}

\author{\bf Sanjukta Chakraborty$^1$,~Subenoy
Chakraborty$^1$\footnote{subenoyc@yahoo.co.in} and Ujjal
Debnath$^2$\footnote{ujjaldebnath@yahoo.com}}

\affiliation{$^1$Department of Mathematics, Jadavpur University,
Calcutta-32, India.\\ $^2$Department of Mathematics, Bengal
Engineering and Science University, Shibpur, Howrah-711 103,
India.\\}

\date{\today}

\begin{abstract}
We study gravitational collapse in higher dimensional
quasi-spherical Szekeres space-time  for matter with anisotropic
pressure. Both local and global visibility of central curvature
singularity has been studied and it is found that with proper
choice of initial data it is possible to show the validity of CCC
for six and higher dimensions. Also the role of pressure in the
collapsing process has been discussed.
\end{abstract}

\pacs{04.20, 04.20 Dw, 04.70 B}

\maketitle

\section{\normalsize\bf{Introduction}}

Usually, for cosmological phenomena over galactic scale or in the
smaller scale, it is reasonable to consider inhomogeneous
solutions to Einstein equations. Szekeres' [1] in 1975, gave a
class of inhomogeneous solutions representing irrotational dust.
The space-time represented by these solutions has no killing
vectors and it has invariant family of spherical hypersurfaces.
Hence this space-time is referred as quasi-spherical space-time.
Recently, Chakraborty et al [2] have extended the Szekeres
solution to $(n+2)$ dimensional space-time and generalized it
for matter containing heat flux [3].\\

In classical general relativity, one of the challenging issues is
gravitational collapse. This problem became important after the
formulation of famous singularity theorems [4] and Cosmic
Censorship Conjecture (CCC)[5]. Also in the perspective of black
hole physics and its astrophysical implications, the end state of
collapse (black hole or naked singularity) is interesting. As
there exists no formal method to address this problem so it is
natural to study various examples of collapsing system (namely,
Tolman-Bondi-Lema\^{\i}tre (TBL) spherically symmetric model
[6-15] or quasi-spherical Szekeres' model [16-18]) with a view to
gain some insight. In general, these studies conclude that the
local or global visibility of the central curvature singularity
depends on the initial data.\\

The above studies are mostly confined to the dust model, there are
very few works on anisotropic stress [19] (confined to TBL model).
For the last few years, there are attempts to study collapse
dynamics in TBL model with anisotropic pressure to address the
question ``can non-zero pressures within a collapsing matter cloud
avoid a naked singularity forming as the end state of a continual
gravitational collapse?'' But so far, the actual role the
pressures play in determining the end state of collapse is not yet
clearly understood. Due to complicated nature of the space-time
geometry there are no works on gravitational collapse with
anisotropic stresses in quasi-spherical model except recently by
Chakraborty et al [20] where they have shown the role of pressure
in 4 dimension.\\

In this work, we extend this study to $(n+2)$ dimensional
Szekeres' model and examine the role of the dimension on collapse
dynamics. The paper is organized as follows: Higher dimensional
Szekeres' model is described in section II, while collapse
dynamics including the study of geodesic is presented in section
III. In section IV, there are discussion and concluding remarks.
Finally at the end
there are two appendix dealing with detailed calculations.\\

\section{\normalsize\bf{Higher Dimensional Szekeres' Model}}
The metric ansatz for (n+2)dimensional Szekeres' space-time is of
the form

\begin{equation}
ds^{2}=dt^{2}-e^{2\alpha}dr^{2}-e^{2\beta}\sum_{i=1}^{n}dx_{i}^{2}\
\end{equation}

where the metric coefficients $\alpha$ and $\beta$ are functions
of all space-time co-ordinates i.e.,
$$\alpha=\alpha(t,r,x_{1},....,x_{n}),~~ \beta=\beta(t,r,x_{1},....,x_{n}).$$

Now Considering both radial and transverse stresses the energy
momentum tensor has the structure

$$T_{\mu}^{\nu}=\text{diag}(\rho,-p_{r},-p_{_{T}},-p_{_{T}})$$

and the compact form of the  Einstein equations are

 \begin{equation}
 \begin{array}{c}
 \rho=\frac{F'}{\zeta^{n}\zeta'}\\\\
 p_{r}=-\frac{\dot{F}}{\zeta^{n}\dot{\zeta}}\\\\
 p_{_{T}}=p_{r}+\frac{\zeta p_{r}'}{n\zeta'}
 \end{array}
 \end{equation}
 where $F(r,t)=\frac{n}{2}{R^{n-1}e^{(n+1)\nu}(\dot{R}^{2}-f(r))}$ and $\zeta=e^{\beta}$.\\

Further, the expressions for the metric functions are

\begin{equation}
 \begin{array}{c}
e^{\beta}=R(t,r)~e^{\nu(r,x_{1},...,x_{n})}~,\\\\
e^{\alpha}=R'+R~\nu'
 \end{array}
 \end{equation}

and the evolution equation for R gives

\begin{equation}
R\ddot{R}+\frac{1}{2}{(n-1)\dot{R}^{2}}+\frac{p_{r}}{n}{R^{2}}=\frac{n-1}{2}~f(r),
~~~~~(f(r)=\text{arbitrary separation function})
\end{equation}
 Also the function $\nu$ satisfies

 \begin{equation}
 e^{-2\nu}\sum_{i=1}^{n}[{(n-2)\nu_{x_{i}}^{2}+2\nu_{x_{i}x_{i}}}]=n(f(r)-1)
 \end{equation}
 which has a solution of the form
 \begin{equation}
e^{-\nu}=A(r)\sum_{i=1}^{n}x_{i}^{2}+\sum_{i=1}^{n}B_{i}(r)x_{i}+C(r)
 \end{equation}
with the restriction,
 \begin{equation}
\sum_{i=1}^{n}B_{i}^{2}-4AC=f(r)-1
\end{equation}

for the arbitrary functions $A(r)$, $B_{i}(r),(i=1,2,..,n)$ and
C(r).\\

As we are considering quasi-spherical gravitational collapse so it
is natural to assume the initial configuration (from which the
collapse has started) to be smooth everywhere. Thus $p_{r}$ should
be regular initially at the center and blows up at the
singularity. So a natural choice for $p_{r}$ is

 \begin{equation}
p_{r}=\frac{g(r)}{R^{l}}
\end{equation}

where the arbitrary function $g(r)$  has the form $r^{l}$ near
$r=0$ to make initial $p_{r}$ finite (non-zero) at the centre
$r=0$ and $l$ is any constant. Hence, the expressions for matter
density and tangential stress become

\begin{equation}
\rho=\frac{H'+(n+1)H\nu'}{R^{n}(R'+R\nu')}
\end{equation}
and
\begin{equation}
p_{_{T}}=\frac{g(r)}{R^{l}}\left[1-\frac{lR'}{n(R'+R\nu')}\right]+\frac{g'(r)}{nR^{l-1}(R'+R\nu')}
\end{equation}

where
$H(R,t)=D(r)-\frac{g(r)}{(n-l+1)}~R^{n-l+1}~,~~~~(l\ne(n+1))$ and
$D(r)$ , an arbitrary integration function.\\

Now, due to this choice of $p_{r}$ (see eq (8)) the evolution eq
(4) for $R$ can be integrated once and the radial velocity of
collapsing shell at a distance $r$ from the centre is given by

\begin{equation}
\dot{R}^{2}=f(r)+\frac{2H(R,t)}{nR^{n-1}}
\end{equation}

This is termed as the equation of the collapsing process.\\

\section{\normalsize\bf{Collapse Dynamics}}

To characterize the nature of the singularity (black hole or
naked singularity), the event horizon of observers at infinity
plays an important role. But formation of event horizon depends
greatly on the computation of null geodesics whose computation are
almost impracticable for the present space-time geometry. So a
closely related concept of a trapped surface (a space-like
2-surface whose normals on both sides are future pointing
converging null geodesic families) will be considered. Thus, if
the 2-surface $S_{r,t}$ ($r=$constant, $t=$constant) is a trapped
surface then it and its entire future development lie behind the
event horizon provided the density falls off fast enough at
infinity. Hence mathematically, if $K^{\mu}$ denotes the tangent
vector field to the null geodesics which is normal to $S_{r,t}$
then we have
$$
K_{\mu}~K^{\mu}=0,~~K^{\mu}_{~;~\nu}~K^{\nu}=0~.
$$

Now the  null geodesics will converge (or diverge) if the
invariant $K^{\mu}_{~;~\mu}<0$ (or $K^{\mu}_{~;~\mu}>0$) on the
surface $S_{r,t}=0$.\\

As a consequence, it is easy to show that the inward geodesics
converges initially and throughout the collapsing process while
the outward geodesics diverges initially but becomes convergent
after a time $t_{ah}(r)$ (time of formation of apparent horizon)
given by
$$
\dot{R}(t_{ah}(r),r)=-\sqrt{1+f(r)}
$$

Now using equations (8) and (11) we have

\begin{equation}
g(r)R^{n+1-l}(t_{ah}(r),r)+\frac{n}{2}(n+1-l)R^{n-1}(t_{ah}(r),r)-(n+1-l)D(r)=0
\end{equation}

From Appendix II, it is to be noted that the central singularity
(at $r=0$) forms at time $t_{0}$ while a trapped surface is formed
at a distance $r$ at time $t_{ah}(r)$ and their difference is
given by the equation (37). Thus if the trapped surface is formed
at a later instant than $t_{0}$ then it is possible for light
signals from the singularity to  reach a distant observer. Hence,
$t_{ah}(r)>t_{0}$ is the necessary condition for formation of
naked singularity and on the otherhand, $t_{ah}(r)\le t_{0}$ is
the sufficient condition for black hole formation. Also it should
be mentioned that this criterion for naked singularity is purely local.\\

Further, as the time difference equation (37) is complicated, so
to make a comparative study between $t_{ah}$ and $t_{0}$ one can
choose for simplicity $l=(n+1)/2$ and equation (37) takes the form

\begin{equation}
t_{ah}(r)-t_{0}=\frac{\sqrt{2n}\left(D_{0}g_{1}-D_{1}g_{0}-g_{1}\sqrt{D_{0}}\sqrt{D_{0}-
\frac{2}{n+1}g_{0}}\right)}{(n+1)g_{0}\sqrt{D_{0}}\sqrt{D_{0}-\frac{2}{n+1}g_{0}}\left(\sqrt{D_{0}}+
\sqrt{D_{0}-\frac{2}{n+1}g_{0}}\right)}~r+O(r^{2})
\end{equation}

The following table shows the possibility of naked singularity or
a black hole under different conditions:\\

\[
\text {TABLE-I}
\]
\[
\begin{tabular}{|l|l|r|r|r|}
\hline\hline \multicolumn{1}{|c|}{~~~Choice of the parameters~~~}
& \multicolumn{1}{c|}{~~~~Naked Singularity~~~~} &
\multicolumn{1}{c|}{~~~~Black hole~~~~}  \\
\hline\hline
&  &     \\
(i)~~ $g_{1}>0, D_{1}<0$  &  ~~~~~~~~~~Always possible  &  Not possible ~~~~~~~~~~~~\\

\hline
&  &     \\
(ii) ~$g_{1}<0, D_{1}>0$  &  ~~~~~~~~~~Not possible &  Always possible ~~~~~~~~\\

\hline
&  &     \\
(iii) $g_{1}>0, D_{1}>0$  &
$\frac{g_{1}}{D_{1}}>\frac{n+1}{2}\left(1+\sqrt{1-\frac{2}{n+1}~\frac{g_{0}}{D_{0}}}\right)$
&
$\frac{g_{1}}{D_{1}}<\frac{n+1}{2}\left(1+\sqrt{1-\frac{2}{n+1}~\frac{g_{0}}{D_{0}}}\right)$
\\

\hline
&  &    \\
(iV) $g_{1}<0, D_{1}<0$  &
$|\frac{g_{1}}{D_{1}}|<\frac{n+1}{2}\left(1+\sqrt{1-\frac{2}{n+1}~\frac{g_{0}}{D_{0}}}\right)$
&
$|\frac{g_{1}}{D_{1}}|>\frac{n+1}{2}\left(1+\sqrt{1-\frac{2}{n+1}~\frac{g_{0}}{D_{0}}}\right)$
\\

\hline\hline
\end{tabular}%
\]%
\newline

From the table, to make the initial density gradient to be
negative at the centre (i.e., $\rho_{1}<0$) one must have
$(D_{1}-\frac{2g_{1}}{n+1})<0$ (for $\nu_{_{-1}}>-1$). For the
first case (i.e., $g_{1}>0,D_{1}<0$) $\rho_{1}$ is negative
definite and there is always naked singularity as in the dust
model. Similarly, $\rho_{1}$ is positive definite in the second
case which leads to  black hole solution same as dust model. For
the third and fourth cases (when $g_{1}$ and $D_{1}$ have same
sign) either NS or BH is possible depending on the restrictions
given in the table I (see also figs 1 - 6). However, in the last
two cases, for $\rho_{1}>0$ only black hole solution is
possible but for $\rho_{1}<0$, both NS and BH are possible.\\

Further, if we assume that $D_{1}=0=g_{1}$ then the time
difference in eq (37) becomes
\begin{equation}
t_{ah}(r)-t_{0}=\sqrt{\frac{n}{2}}\left[-2^{\frac{3n-1}{2n-2}}
n^{\frac{n+1}{2-2n}}~_{2}F_{1}[\frac{1}{2},b,b+1,z{\left(\frac{2D_{0}}{n}\right)}
^{\frac{n+1-l}{n-1}}]~D_{0}^{\frac{1}{n-1}}~r^{\frac{3n+3-2l}{n-1}}+...
... ...\right]
\end{equation}

We see that if $n>3$ (with $l<n+1$) then the first term on the
right side will be dominating compare to other terms and hence we
always have $t_{ah}<t_{0}$. Thus in this case black hole is the
only final state collapse for six and higher dimensional
space-times. This distinctive result is similar to dust collapse
[13, 16] and we conclude that CCC is valid in this case for six
and higher dimension with anisotropic pressure.\\\

\begin{figure}
\includegraphics[height=1.7in]{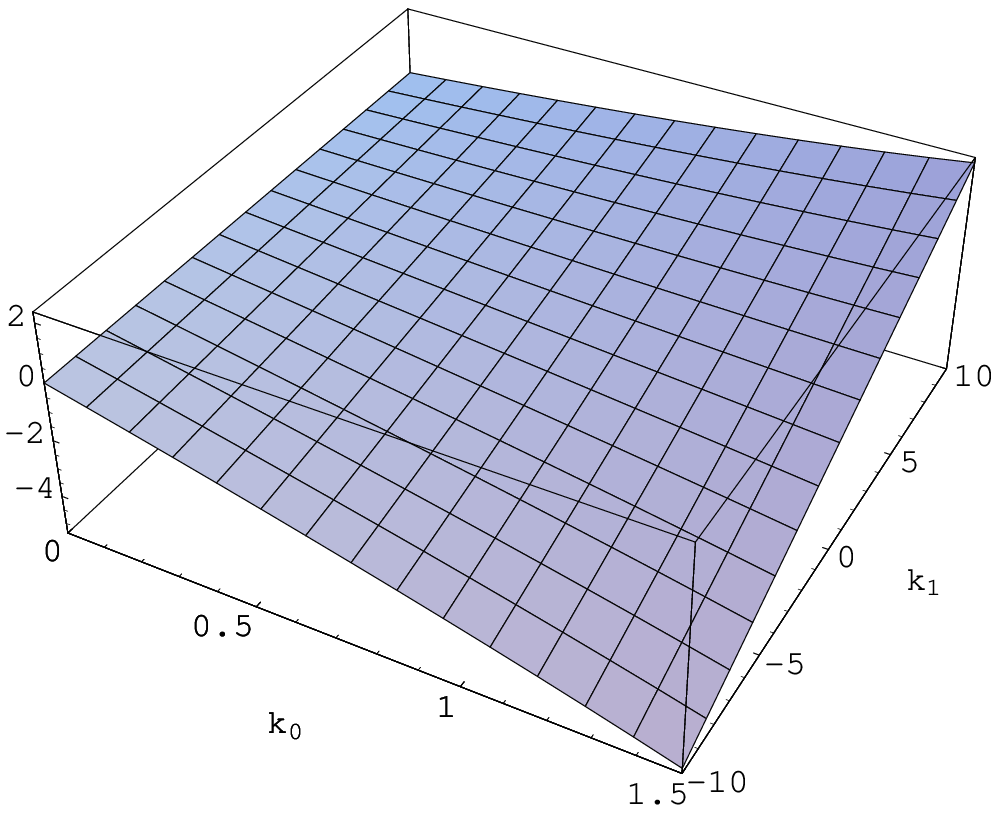}~~~
\includegraphics[height=1.7in]{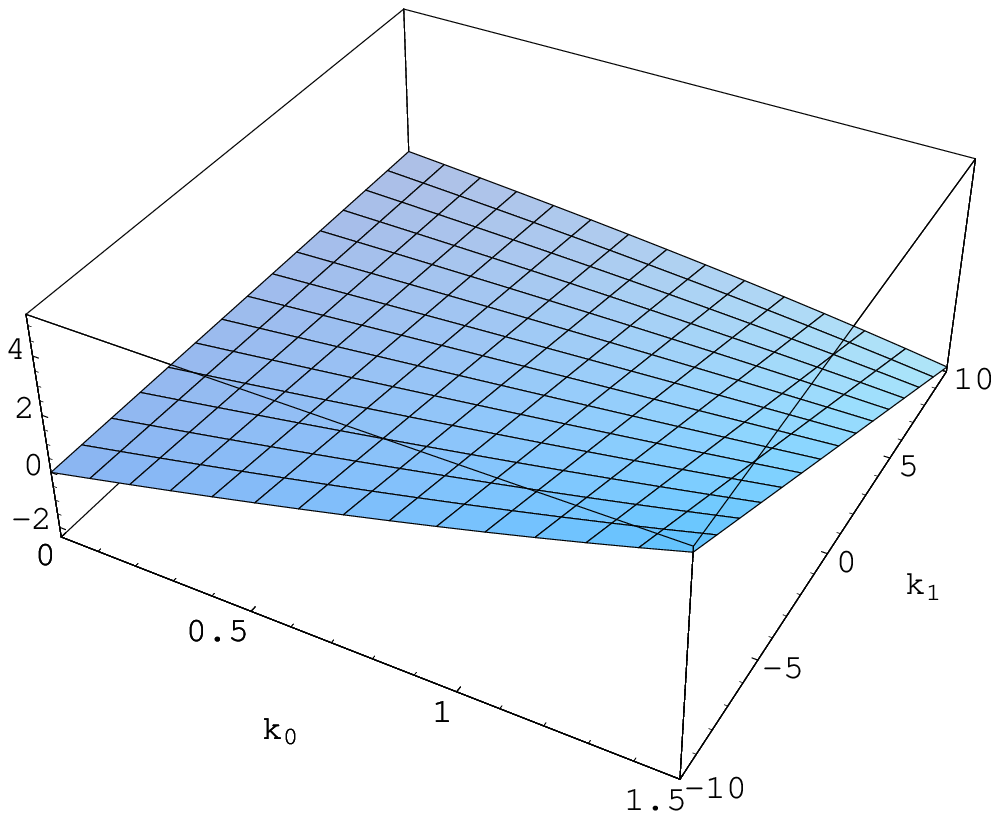}\\
\vspace{1mm}
Fig.1~~~~~~~~~~~~~~~~~~~~~~~~~~~~~~~~~~~~~~~~~~~Fig.2\\
\vspace{5mm} \hspace{1cm} \vspace{6mm}

\includegraphics[height=1.7in]{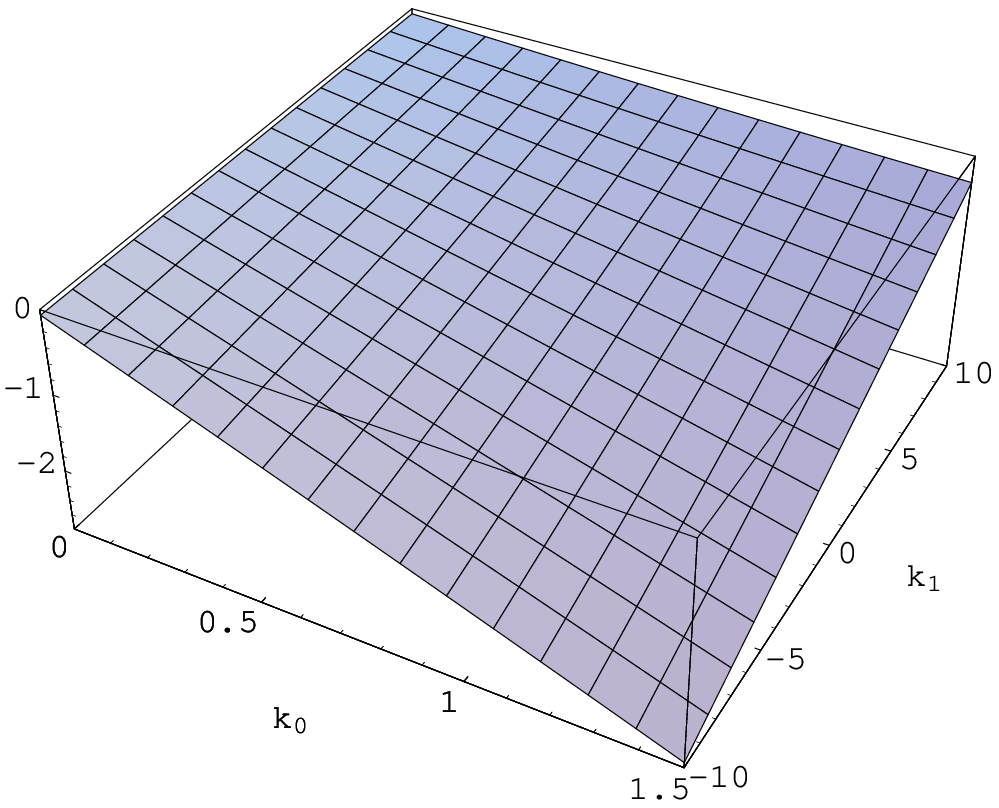}~~~
\includegraphics[height=1.7in]{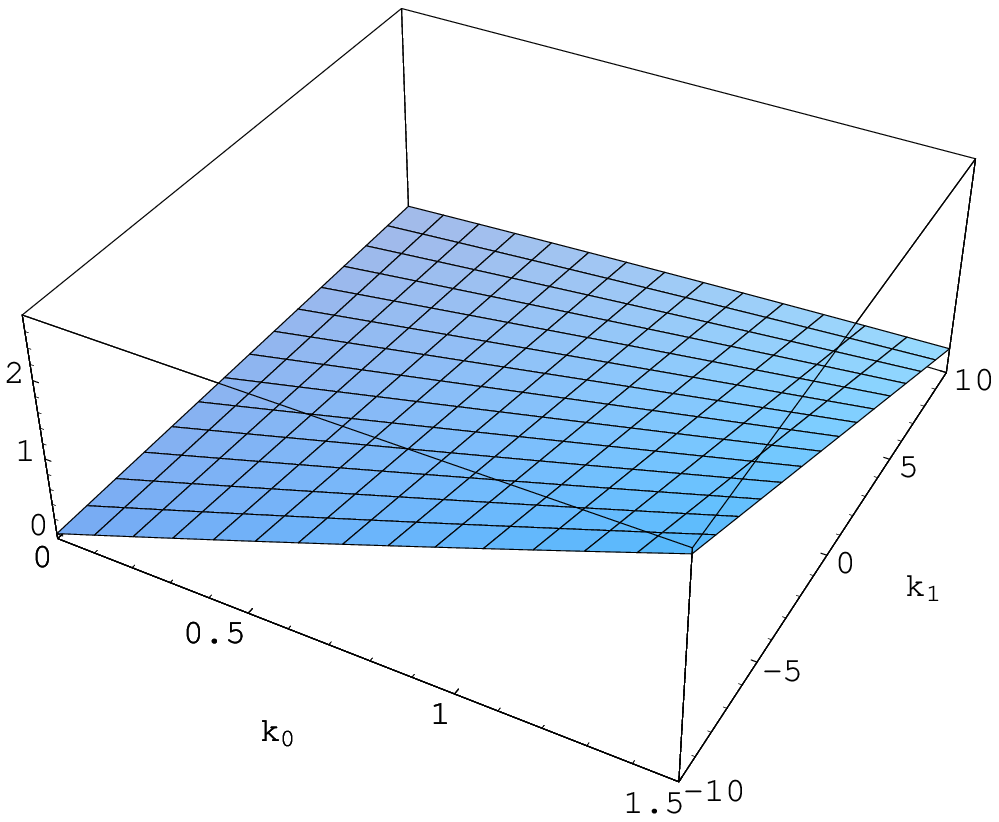}\\
\vspace{1mm}
Fig.3~~~~~~~~~~~~~~~~~~~~~~~~~~~~~~~~~~~~~~~~~~~Fig.4\\
\vspace{5mm}\hspace{1cm} \vspace{6mm}

\includegraphics[height=1.7in]{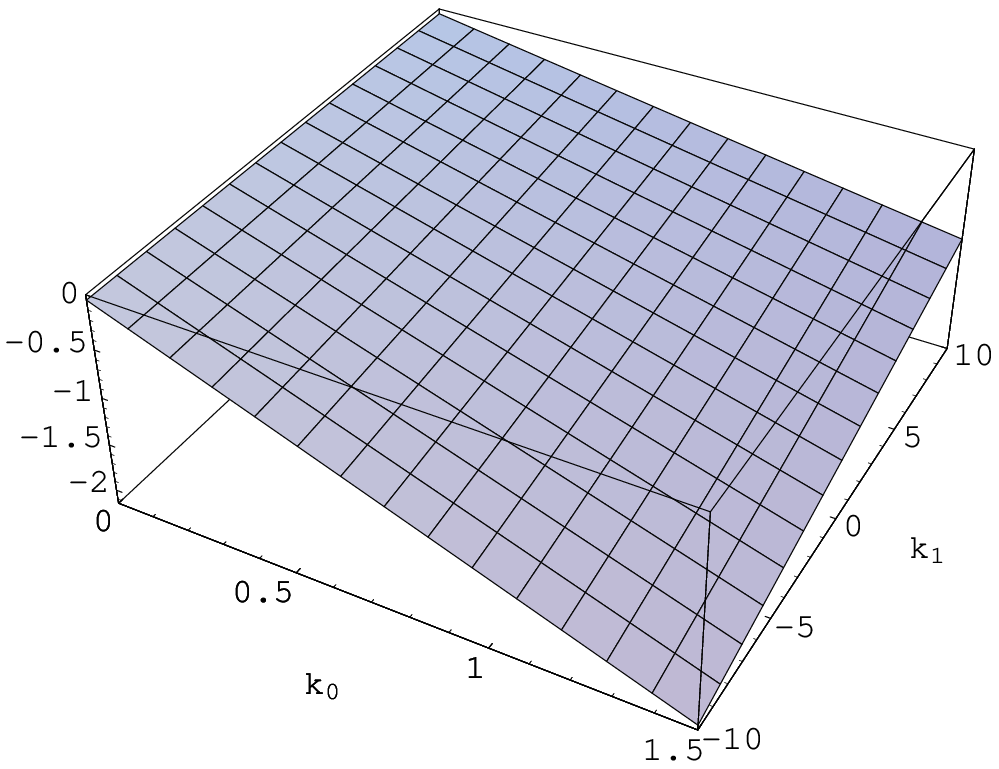}~~~
\includegraphics[height=1.7in]{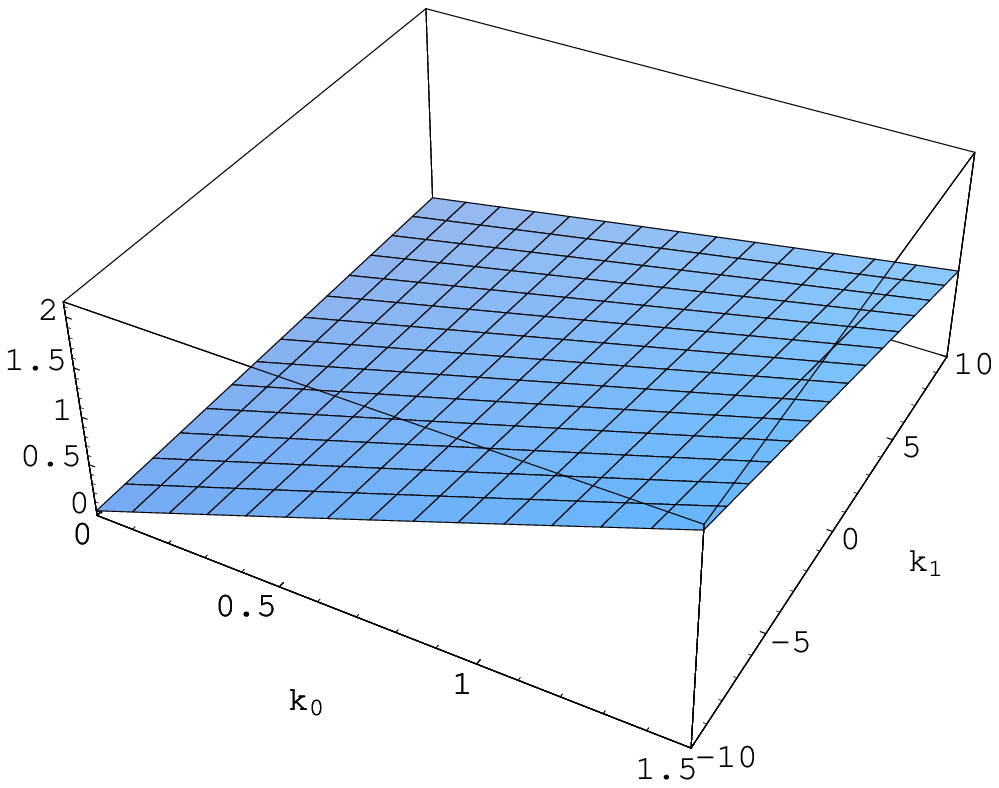}\\
\vspace{1mm}
Fig.5~~~~~~~~~~~~~~~~~~~~~~~~~~~~~~~~~~~~~~~~~~~Fig.6\\
\vspace{5mm} Figs. 1 - 6 show variation of $t_{ah}-t_{0}$ of
eq.(13) for the variation of $k_{0}(=g_{0}/D_{0})$ and
$k_{1}(=g_{1}/D_{1})$. Figs.1, 3 and 5 correspond to $D_{1}>0$ for
$n=4, 12$ and 25 respectively while Figs.2, 4  and 6 correspond to
$D_{1}<0$ for $n=4, 12$ and 25 respectively. \hspace{1cm}
\vspace{6mm}
\end{figure}

\subsection{\normalsize\bf{Study of Geodesics}}

In this section, the nature of the singularity (NS or BH) is
examined by studying the geodesics from the singularity. In
particular, it will be investigated whether there exist one or
more radial outgoing null geodesics which terminate in the past at
the central singularity.  For simplicity of calculation only
marginally bound case ($f(r)=0$) with $l=(n+1)/2$ will be
considered (as in the previous section). Now choosing the initial
time $t_{i}=0$, the explicit solution for $R(t,r)$  can be written
as

\begin{equation}
t(r)=\frac{\sqrt{2n}}{g(r)}\left[\sqrt{D(r)-\frac{2}{n+1}g(r)R^{(n+1)/2}}-\sqrt{D(r)-\frac{2}{n+1}g(r)r^{(n+1)/2}}\right]
\end{equation}

Thus the expression for the singularity time for the shell of
radius $R$ is given by ($R(t_{s}(r),r)=0$)

\begin{equation}
t_{s}(r)=\frac{\sqrt{2n}}{g(r)}\left[\sqrt{D(r)}-\sqrt{D(r)-\frac{2}{n+1}g(r)r^{(n+1)/2}}\right]
\end{equation}

and consequently the time for central singularity is

\begin{equation}
t_{0}=\frac{\sqrt{2n}}{g_{0}}\left(\sqrt{D_{0}}-\sqrt{D_{0}-\frac{2}{n+1}g_{0}}\right)
\end{equation}

Here the polynomial form of $D(r)$ and $g(r)$ are taken in the
form

\begin{equation}\begin{array}{c}
D(r)=D_{0}r^{n+1}+D_{k}r^{k+n+1}
\\\\g(r)=g_{0}~r^{(n+1)/2}+g_{_{j}}~r^{j+(n+1)/2}
\end{array}
\end{equation}

where $D_{0}, g_{0}$ are constants and $D_{k} (<0)$ and $g_{_{j}}
(<0)$ are the first non-vanishing term beyond $D_{0}$ and $g_{0}$
respectively. Now using these expression for $D(r)$ and $g(r)$ the
time of collapse of a typical shell of radius $R$ (i.e.,
$t_{s}(r)$) becomes (see eq (16))

\begin{equation}
t_{s}(r)=t_{0}+\sqrt{\frac{n}{2}}\frac{D_{k}}{g_{0}}\left(\frac{1}{\sqrt{D_{0}}}-\frac{1}{\sqrt{D_{0}-
\frac{2}{n+1}g_{0}}}\right)r^{k}+\frac{\sqrt{2n}~g_{_{j
}}}{g_{0}}\left(\frac{1}{(n+1)\sqrt{D_{0}-\frac{2}{n+1}g_{0}}}-
\frac{\sqrt{D_{0}}-\sqrt{D_{0}-\frac{2}{n+1}g_{0}}}{g_{0}}
\right)r^{j}+....
\end{equation}

Now the equation of the outgoing radial null geodesic (ORNG) which
passes through the central singularity, can be chosen to be (near
$r=0$)

\begin{equation}
t_{ORNG}=t_{0}+a~r^{\xi}
\end{equation}

where $a(>0)$ and $\xi(>0)$ are constants.\\

In the polynomial form for $D(r)$ and $g(r)$ in the equation (18)
one can choose two possibilities:
$$(i)~~k<j~,~~~~~~(ii)~~k>j$$

{\bf Case I} :~~  $k<j$ : \\

Here near $r=0$, the expression for $t_{s}(r)$ can be written as
(see eq.(19))
\begin{equation}
t_{s}(r)=t_{0}-\sqrt{\frac{n}{2}}\frac{D_{k}}{g_{0}}\left(\frac{1}{\sqrt{D_{0}-\frac{2}{n+1}g_{0}}}-\frac{1}{\sqrt{D_{0}}}
\right)r^{k},~~~~(D_{k}<0)
\end{equation}

In order that null geodesic passes through the shell of radius $R$
before the trapped surface is formed there, comparing (20) and
(21) one gets  ($t_{ORNG}<t_{s}(r)$)
\begin{equation}
(a)~~\xi>
k~~\text{or}~~~(b)~~\xi=k~~~\text{and}~~~a<-\frac{D_{k}}{g_{0}}\sqrt{\frac{n}{2}}\left(\frac{1}{\sqrt{D_{0}-\frac{2}{n+1}g_{0}}}-\frac{1}{\sqrt{D_{0}}}
\right)
\end{equation}

When $\xi>k$ then near $r=0$ the solution for $R$ simplifies to

\begin{equation}
R=r\left[1-\frac{n+1}{4n}g_{0}t^{2}-\frac{n+1}{\sqrt{2n}}~t\left(\sqrt{D_{0}-\frac{2}{n+1}g_{0}}+\frac{D_{k}r^{k}}{2\sqrt{D_{0}-
\frac{2}{n+1}g_{0}}}\right) \right]^{2/n+1}
\end{equation}

Further for the given metric an ORNG should satisfy

\begin{equation}
\frac{dt}{dr}=R'+R\nu'
\end{equation}

Now using (20) and (23) in  (24) one gets (up to leading order in
$r$)

\begin{equation}
a\xi
r^{\xi-1}=\left(1+\nu_{_{-1}}+\frac{2k}{n+1}\right)\left[-\frac{(n+1)D_{k}t_{0}}{2\sqrt{2n}\sqrt{D_{0}-
\frac{2}{n+1}g_{0}}}\right]^{2/(n+1)}r^{\frac{2k}{n+1}}
~,~~~~~(\nu_{_{-1}}\ne 0)
\end{equation}

This gives
\begin{equation}
\xi=1+\frac{2k}{n+1}>0
~~~\text{and}~~~a=\frac{1}{\xi}\left(1+\nu_{_{-1}}+\frac{2k}{n+1}\right)\left[-\frac{(n+1)D_{k}t_{0}}
{2\sqrt{2n}~\sqrt{D_{0}-\frac{2}{n+1}g_{0}}}\right]^{2/(n+1)}
\end{equation}

As $\xi>k$, so from the above relations (26)

\begin{equation}
\begin{array}{c}
k<\frac{n+1}{n-1}~~ \text{and} ~~ \xi<\frac{n+1}{n-1},\\\\
\text{i.e. one could have ($n>2$)}\\\\
 k=1,~\xi=\frac{n+3}{n+1}
\end{array}
\end{equation}

On the other hand for $\xi=k$, as before $k=\frac{n+1}{n-1}$ and

\begin{eqnarray*}
a=\frac{n-1}{n+1}\left[-\frac{n+1}{4}\left(\frac{2ag_{0}t_{0}}{n}+2a\sqrt{\frac{2}{n}}~\sqrt{D_{0}-\frac{2}{n+1}g_{0}}+
\frac{D_{k}t_{0}\sqrt{\frac{2}{n}}}{\sqrt{D_{0}-\frac{2}{n+1}g_{0}}}\right)\right]^{(1-n)/(n+1)}~\times
\end{eqnarray*}

\begin{equation}
\left[-\frac{n+1}{4}
\left\{(1+\nu_{_{-1}})\left(\frac{2ag_{0}t_{0}}{n}+2a\sqrt{\frac{2}{n}}~\sqrt{D_{0}-\frac{2}{n+1}g_{0}}\right)+
\frac{\left(1+\nu_{_{-1}}+\frac{2k}{n+1}\right)~\sqrt{\frac{2}{n}}D_{k}t_{0}}{\sqrt{D_{0}-\frac{2}{n+1}g_{0}}}
\right\} \right]
\end{equation}

Thus for $\xi>k$, $k$ has only one value (namely $k=1$) and $n$
can take any value ($>2$) while for $\xi=k$ the only possible
values of $n$ are 2 and 3 only. Therefore, geodesic equations are
possible in any dimension for $\xi>k$ but it is only possible up
to five dimension for $\xi=k$. In other words, for $\xi=k$, naked
singularity is possible only up to five dimension which supports
the results in the previous section. Lastly, it should be
mentioned that the other choice namely $k>j$ is similar to the
above and hence not presented here .\\

\section{\normalsize\bf{Discussions and Concluding Remarks}}

A detailed analysis of the central curvature singularity as the
final state of collapse in the ($n+2$)-dimensional quasi-spherical
Szekeres' model has been done for matter with anisotropic pressure
(i.e., both radial and tangential pressures are non-zero and
distinct). The local visibility of the central singularity has
been discussed by comparing the time of formation of trapped
surface and the time of formation of central shell focusing
singularity while global visibility is examined by considering
only the radial null geodesics (for simplicity). Most of the
results are very similar to that for four dimension in ref. [13].
It is to be noted that though we have considered quasi-spherical
Szekeres model but still these are valid for TBL model. In fact,
in Szekeres' solution if we assume the function $\nu$ to be
independent of $r$ (i.e., $\nu'=0$) then Szekeres' model can be
converted to TBL model by the following coordinate transformation:\\

\[
\begin{array}{llll}
x_{1}=Sin\theta _{n}Sin\theta _{n-1}...~~...Sin\theta _{2}Cot\frac{1}{2}%
\theta _{1} &  &  &  \\
&  &  &  \\
x_{2}=Cos\theta _{n}Sin\theta _{n-1}...~~...Sin\theta _{2}Cot\frac{1}{2}%
\theta _{1} &  &  &  \\
&  &  &  \\
x_{3}=Cos\theta _{n-1}Sin\theta _{n-2}...~~...Sin\theta _{2}Cot\frac{1}{2}%
\theta _{1} &  &  &  \\
&  &  &  \\
....~~...~~...~~...~~...~~...~~...~~... &  &  &  \\
&  &  &  \\
x_{n-1}=Cos\theta _{3}Sin\theta _{2}Cot\frac{1}{2}\theta _{1} &  &  &  \\
&  &  &  \\
x_{n}=Cos\theta _{2}Cot\frac{1}{2}\theta _{1} &  &  &
\end{array}%
\]%
\newline

Further, the radial pressure is assumed to be a function of `$r$'
and `$t$' only with the form  $p_{r}=g(r)/R^{l}$. Throughout the
paper $l$ is chosen to be less than $n+1$ (i.e., $l<n+1$)  [It is
to be noted that for $l=n+1$ due to appearance of a logarithmic
term the calculations become very much complicated while for
$l>n+1$ the results are not of much interest]. It has been shown
that for $g_{1}=0=D_{1}$ i.e., for $h_{1}=0$, naked singularity is
only possible up to five dimension---a result identical in dust
collapse. The above choice (i.e., $h_{1}=0$) gives
$\rho_{0}=(n+1)h_{0}$ for $\nu_{-1}\ge(-1)$ and $\rho_{1}=0$ for
$\nu_{-1}>-1$. Hence for $\nu_{-1}>-1$, if the initial density
gradient falls off at the centre ($r=0$) then one can say that six
dimension plays as critical dimension for naked singularity, a
distinct result in higher dimension. Further with the choice
$l=(n+1)/2$, a detailed comparative study has been done between
the time of formation of trapped surface and that of central
singularity and Table I shows all possibilities for the parameters
involved in the expression. Also in the figures 1 - 6 we have
shown graphically the time difference $t_{ah}-t_{0}$
for 6, 14 and 27 dimensions for $D_{1}>0$ and $D_{1}<0$ respectively.\\

As in dust collapse,in this case we have definitely a black hole
(or naked singularity) if the initial density gradient at the
centre is positive definite (or negative definite). But in the
indefiniteness in the sign of $\rho_{1}$ one may note that if the
initial density and radial pressure has identical behaviour (i.e.,
increase or decrease simultaneously) then even with initial
negative density gradient (at the centre) it is possible to have
black hole as the end state of collapse, while if the initial
density and pressure have opposite nature (i.e., one increase when
other decreases and vice versa) then the behaviour is identical to
dust collapse. Therefore one may conclude that pressure tries to
resist the formation of naked singularity.

$$
\text{\bf APPENDIX ~I}
$$

{\bf Initial hypersurface and the physical parameters:}\\

Suppose the collapsing process starts on the initial hypersurface
$(t=t_{i})$ and we have $R=r$ there. Then the expressions for
energy density, radial pressure and the tangential pressure, at
the beginning of the collapse are

\begin{equation}
\begin{array}{c}
\rho_{i}(r,x_{1},...,x_{n})=\rho(t_{i},r,x_{1},...,x_{n})=\frac{h'(r)+(n+1)h(r)\nu'}{r^{n}(1+r\nu')}\\\\

p_{_{T_{i}}}(r,x_{1},...,x_{n})=p_{_{T}}(t=t_{i})=\frac{g(r)}{r^{l}}\left[1-\frac{l}{n(1+r\nu')}\right]+\frac{g'(r)}{nr^{l-1}(1+r\nu')}
\end{array}
\end{equation}

where $h(r)=H(r,t_{i})=D(r)-\frac{g(r)}{n-l+1}~r^{n-l+1}$.\\

For smooth initial data  $h(r)$ and $g(r)$ to be $C^{\infty}$
functions and hence one can choose the following series expansions

\begin{equation}
\begin{array}{c}
D(r)=\sum_{j=0}^{\infty}D_{j}~r^{n+1+j}\\\\
g(r)=\sum_{j=0}^{\infty}g_{j}~r^{l+j}\\\\
\rho_{i}(r,x_{1},...,x_{n})=\sum_{j=0}^{\infty}\rho_{j}~r^{j}\\\\
\nu'(r,x_{1},...,x_{n})=\sum_{j=-1}^{\infty}\nu_{j}~r^{j}~,~~~~~~~(\nu_{_{-1}}\ge
-1).\\\\
p_{_{T_{i}}}(r,x_{1},...,x_{n})=\sum_{j=0}^{\infty}p_{j}~r^{j}
\end{array}
\end{equation}

It is to be noted that in the above series expansions the
coefficients  $D_{j}$'s and $g_{j}$'s ($j=0, 1, ...$) are purely
constants while $\rho_{j}$'s, $\nu_{j}$'s and $p_{j}$'s are
functions of $x_{i}$'s ($i=1,...,n$). Also these coefficients are
not independent but are  related among themselves through the
relations in eq (29)  as follows:\\

For $\nu_{_{-1}}>-1$:
\begin{equation}\begin{array}{c}
p_{0}=g_{0},~~
p_{1}=g_{1}\left\{1+\frac{1}{n(1+\nu_{_{-1}})}\right\},~~p_{2}=g_{2}
\left\{1+\frac{1}{(1+\nu_{_{-1}})}\right\}-\frac{g_{1}~\nu_{0}}{n(1+\nu_{_{-1}})^{2}},~~....~~....\\\\
\rho_{0}=(n+1)h_{0},~~\rho_{1}=\frac{(n+2)+(n+1)\nu_{_{-1}}}{1+\nu_{_{-1}}}~h_{1},~~\rho_{2}=
\frac{(n+3)+(n+1)\nu_{_{-1}}}{1+\nu_{_{-1}}}~h_{2}-\frac{\nu_{0}~h_{1}}{(1+\nu_{_{-1}})^{2}},~~....~~....
\end{array}
\end{equation}

For $\nu_{_{-1}}=-1$:
\begin{equation}\begin{array}{c}
p_{0}=g_{0}+\frac{g_{1}}{n\nu_{0}},~~
p_{1}=g_{1}\left(1-\frac{\nu_{1}}{n\nu_{0}^{2}}\right)+\frac{2g_{2}}{n\nu_{0}},~~p_{2}=g_{2}
\left(1-2\frac{\nu_{1}}{n\nu_{0}^{2}}\right)+\frac{(\nu_{1}^{2}-\nu_{0}\nu_{2})}{n\nu_{0}^{3}}~g_{1}+
\frac{3g_{3}}{n\nu_{0}},~~....~~....\\\\
\rho_{0}=(n+1)h_{0}+\frac{h_{1}}{\nu_{0}},~~\rho_{1}=\frac{2h_{2}}{\nu_{0}}+h_{1}\left((n+1)-\frac{\nu_{1}}
{\nu_{0}^{2}}\right),~~\rho_{2}=\frac{3h_{3}}{\nu_{0}}+h_{2}\left((n+1)-\frac{2\nu_{1}}
{\nu_{0}^{2}}\right)+h_{1}\frac{(\nu_{1}^{2}-\nu_{0}\nu_{2})}{\nu_{0}^{3}}
,~~....~~....
\end{array}
\end{equation}
with $h_{i}=D_{i}-\frac{g_{i}}{n-l+1},~~i=0,1,2,...$.\\

$$
\text{\bf APPENDIX ~II}
$$

{\bf Solution of the evolution equation (11) with $f(r)=0$:}\\

For the initial choice $R=r$ at $t=t_{i}$, the explicit solution
is
\begin{equation}
t-t_{i}=-\frac{\sqrt{2n}~r^{\frac{n+1}{2}}}{(n+1)\sqrt{D(r)}}~_{2}F_{1}[\frac{1}{2},b,b+1,
\frac{g(r)r^{n+1-l}}{D(r)(n+1-l)}]-\frac{\sqrt{2n}~R^{\frac{n+1}{2}}}{(n+1)\sqrt{D(r)}}~_{2}F_{1}[\frac{1}{2},b,b+1,
\frac{g(r)R^{n+1-l}}{D(r)(n+1-l)}]
\end{equation}

where $b=\frac{n+1}{2-2l_2n}$ and $_{2}F_{1}$ is the usual
hypergeometric function and $l\ne n+1$.\\

If $t=t_{s}(r)$ stands for time of collapse of the $r$-th shell
i.e., $R(t_{s}(r),r)=0$ then we have

\begin{equation}
t_{s}(r)-t_{i}=\frac{\sqrt{2n}~r^{(n+1)/2}}{(n+1)\sqrt{D(r)}}~_{2}F_{1}[\frac{1}{2},b,b+1,
\frac{g(r)r^{n+1-l}}{D(r)(n+1-l)}]
\end{equation}

Note that $t=t_{s}(r)$ is a monotonic increasing function of $r$
(i.e., $t_{s}'(r)\ge 0$) for the shell focusing singularity. Using
equation (12)

\begin{equation}
t_{ah}(r)-t_{i}=\frac{\sqrt{2n}~r^{(n+1)/2}}{(n+1)\sqrt{D(r)}}~_{2}F_{1}[\frac{1}{2},b,b+1,\frac{g(r)r^{n+1-l}}{D(r)(n+1-l)}]-
\frac{\sqrt{2n}R^{(n+1)/2}(t_{ah},r)}{(n+1)\sqrt{D(r)}}~_{2}F_{1}[\frac{1}{2},b,b+1,\frac{g(r)R^{n+1-l}(t_{ah},r)}{D(r)(n+1-l)}]
\end{equation}

A comparison of equations (34) and (35) shows that the shell
focusing singularity that appears at $r>0$ is in the future of the
apparent horizon. However, the time of occurrence of central shell
focusing singularity (which is of main interest here) is given by

\begin{equation}\begin{array}{c}
\hspace{-6cm}t_{0}~=lim~~~~t_{s}(r)\\
\hspace{-6.3cm}~r\rightarrow 0\\\\
\hspace{.8cm}=t_{i}+\frac{\sqrt{2n}}{(n+1)\sqrt{D_{0}}}~_{2}F_{1}[\frac{1}{2},b,b+1,z],~~~~~\left(z=\frac{g_{0}}{D_{0}(n+1-l)}\right)
\end{array}
\end{equation}

where the series form of $g(r)$  and $D(r)$ (from eq. (30)) have
been used in evaluating the limit. Thus for the restriction
$l<n+1$ , the explicit form of the difference  between $t_{ah}(r)$
and $t_{0}$ is

\begin{eqnarray*}
t_{ah}(r)-t_{0}=\sqrt{\frac{n}{2}}\left[\left\{-\frac{D_{0}^{-3/2}D_{1}}{n+1}~_{2}F_{1}[\frac{1}{2},b,b+1,z]+
\frac{(D_{0}g_{1}-D_{1}g_{0})}{(n+1-l)(3+3n-2l)}D_{0}^{-5/2}~_{2}F_{1}[\frac{3}{2},b+1,b+2,z]r\right\}\right.
\end{eqnarray*}

\begin{equation}
\left.+O(r^{2})-\frac{{2^{\frac{1-3n}{2-2n}}}{n^{\frac{n+1}{2-2n}}}D_{0}^{\frac{1}{n-1}}}
{n+1}~_{2}F_{1}[\frac{1}{2},b,b+1,z{\left(\frac{2D_{0}}{n}\right)}^{\frac{n+1-l}{n-1}}]~r^{\frac{3n+3-2l}{n-1}}+....~~....~\right]
\end{equation}

{\bf Acknowledgement:}\\

One of the authors (SC) is thankful to CSIR, Govt. of India for
providing a research project No. 25(0141)/05/EMR-II.\\

{\bf References:}\\
\\
$[1]$ P. Szekeres, {\it Commun. Math. Phys.} {\bf 41} 55 (1975).\\
$[2]$ S. Chakraborty and U. Debnath, {\it Int. J. Mod. Phys. D} {\bf 13} 1085 (2004)).\\
$[3]$ U. Debnath, S. Nath and S. Chakraborty, {\it Gen. Rel. Grav.} {\bf 37} 215 (2005 ).\\
$[4]$ S.W. Hawking, and G.F.R. Ellis, \textit{''The Large Scale
Structure of Space-Time''.}(Cambridge University Press,
Cambridge, England. 1973).\\
$[5]$ R. Penrose, \textit{Riv. Nuovo Cim.} \textbf{1} 252 (1969);
R.Penrose, \textit{In General Relativity, an Einstein Centenary
Volume, S.W. Hawking and W. Israel (Eds.),} (Camb.
Univ.Press, Cambridge, 1979).\\
$[6]$ P. S. Joshi and I. H. Dwivedi, \textit{Commun. Math. Phys.} \textbf{166%
} 117 (1994).\newline
$[7]$ P. S. Joshi and I. H. Dwivedi, \textit{Class. Quantum Grav.} \textbf{16%
} 41 (1999).\newline
$[8]$ K. Lake, \textit{Phys. Rev. Lett.}
\textbf{68} 3129 (1992).\newline
$[9]$ A. Ori and T. Piran, \textit{Phys. Rev. Lett.} \textbf{59} 2137 (1987).%
\newline
$[10]$ T. Harada, \textit{Phys. Rev. D} \textbf{58} 104015
(1998).\newline
$[11]$ P.S. Joshi, \textit{Global Aspects in Gravitation and Cosmology, }%
(Oxford Univ. Press, Oxford, 1993).\newline
$[12]$ U. Debnath and S. Chakraborty, {\it Gen. Rel. Grav.} {\bf 36} 1243 (2004).\\
$[13]$ A. Banerjee, U. Debnath and S. Chakraborty, \textit{Int. J.
Mod. Phys. D} {\bf 12} 1255 (2003).\\
$[14]$ R. Goswami and P.S. Joshi, \textit{gr-qc}/02112097 (2002)
.\newline $[15]$ S. Schoen and S. T. Yau, \textit{Commun. Math.
Phys.} \textbf{90} 575 (1983).\newline
$[16]$ U. Debnath, S.
Chakraborty and J. D. Barrow, {\it Gen. Rel.
Grav.} {\bf 36} 231 (2004).\\
$[17]$ U. Debnath and S. Chakraborty, {\it JCAP}~ {\bf 05} 001 (2004).\\
$[18]$ R. Goswami and P. S. Joshi, {\it Class. Quantum Grav.} {\bf
21} 3645 (2004); {\it Class. Quantum Grav.} {\bf 19} 5229 (2002).\\
$[19]$ J. R. Gair, {\it Class. Quantum Grav.} {\bf 18} 4897
(2001); R. Goswami and P. S. Joshi, {\it Class. Quantum Grav.}
{\bf 19} 5229 (2002); P. S. Joshi, R. Goswami and N. Dadhich,
{\it Phys. Rev. D}
{\bf 70} 087502 (2004) .\\
$[20]$ S. Chakraborty, S. Chakraborty and U. Debnath, {\it Int. J.
Mod. Phys. D (In press)} (2005); {\it gr-qc}/0506048.\\

\end{document}